# Principal Component Analysis Applied to Gradient Fields in Band Gap Optimization Problems for Metamaterials

Giorgio Gnecco[1], Andrea Bacigalupo[2], Francesca Fantoni[3], and Daniela Selvi[4]

[1]IMT School for Advanced Studies – AXES – Lucca, Italy
[2]University of Genoa – DICCA – Genoa, Italy
[3]University of Brescia – DICATAM – Brescia, Italy
[4]University of Florence – DIEF – Florence, Italy

[1]Corresponding author: giorgio.gnecco@imtlucca.it
[2]andrea.bacigalupo@unige.it
[3]francesca.fantoni@unibs.it
[4]daniela.selvi@unifi.it

**Abstract.** A promising technique for the spectral design of acoustic metamaterials is based on the formulation of suitable constrained nonlinear optimization problems. Unfortunately, the straightforward application of classical gradient-based iterative optimization algorithms to the numerical solution of such problems is typically highly demanding, due to the complexity of the underlying physical models. Nevertheless, supervised machine learning techniques can reduce such a computational effort, e.g., by replacing the original objective functions of such optimization problems with more-easily computable approximations. In this framework, the present article describes the application of a related unsupervised machine learning technique, namely, principal component analysis, to approximate the gradient of the objective function of a band gap optimization problem for an acoustic metamaterial, with the aim of making the successive application of a gradient-based iterative optimization algorithm faster. Numerical results show the effectiveness of the proposed method.

## 1. INTRODUCTION

The analysis and optimal design of acoustic metamaterials is of high interest to the scientific community, since such metamaterials are capable to achieve physical properties hardly found in nature, via a suitable optimization of their geometrical/mechanical model parameters. Application fields include mechanical, civil, naval, and aerospace engineering, up to biomedical and robotic engineering. Unfortunately, due to the complexity of their physical models, the analysis of acoustic metamaterials is often time-consuming, and their optimization cannot be typically obtained neither by hand, nor by applying classical optimization methods, unless in the case in which their models are relatively simple [1]. As demonstrated by recent works (see, e.g., [2, 3]), however, machine learning can make the analysis and optimization of acoustic metamaterials easier, by speeding up their technological transfer process. Moreover, due to the similarity in the respective optimization problems, machine learning techniques can be applied not only for an improved design of acoustic metamaterials, but also for the one of optical metamaterials.

In [2, 3], supervised machine learning techniques were employed to replace the objective function of a band gap optimization problem associated with an acoustic metamaterial by a suitable surrogate objective function (more specifically, by a radial basis function network [4]), which was constructed by using, respectively, a fixed/adaptively chosen interpolating training set. Then, a gradient-based iterative optimization algorithm was applied to optimize the surrogate objective function, by replacing each expensive evaluation of the original objective function by the cheaper computation of its approximation.

In this work, a related but different approach is used, based on unsupervised learning, and more specifically, on a well-known dimensionality reduction technique, Principal Component Analysis (PCA). This allows one to approximate a set of vectors belonging to a Euclidean space by projecting such vectors onto an automatically chosen lower-dimensional subspace, whose dimension is selected in such a way to achieve a desired precision in the approximation [5]. Differently from its standard application to the input vector of a machine learning problem, in the present work PCA is used to approximate the sampled gradient field of a band gap optimization problem of an acoustic metamaterial, evaluated on a subset of admissible vectors of parameters, generated by a subsequence of a quasi-Monte Carlo sequence [6]. Then, a gradient-based iterative optimization algorithm is applied to optimize the objective function, by approximating, at each iteration, the gradient of the objective function with its projection onto the subspace previously obtained by PCA. Because of that projection, the time needed for the execution of each step of the gradient-based iterative optimization algorithm is reduced, while the error in the approximation of the gradient turns out to be negligible due to the automatic choice of the dimension of the subspace. Numerical results show the effectiveness of the proposed computational method of gradient approximation.

## 2. BAND GAP OPTIMIZATION PROBLEM AND DESCRIPTION OF THE PROPOSED METHOD

The following band gap optimization problem for an acoustic metamaterial is considered, assuming continuous differentiability of it objective function. The problem is expressed in the constrained nonlinear programming form

$$\begin{aligned} \underset{\mathbf{x} \in \mathbb{R}^d}{\text{maximize}} \quad & f(\mathbf{x}), \\ \text{s. t.} \quad & g_j(\mathbf{x}) \leq 0, \, j = 1, \ldots, n_{\text{ineq}} \end{aligned} \tag{1}$$

where $\mathbf{x}$ represents a $d$-dimensional Euclidean vector of parameters of the acoustic metamaterial, the nonlinear objective function $f(\mathbf{x})$ models the band gap between two given dispersion surfaces of its Floquet-Bloch spectrum, and $g_j(\mathbf{x})$ are $n_{\text{ineq}}$ (either linear or nonlinear) inequality constraints of either geometrical/mechanical nature, or expressing bounds on the range of each parameter (box constraints). It is worth noting that, in the present work, the band gap $f(\mathbf{x})$ is defined as the difference between the minimum of a dispersion surface and the maximum of the adjacent dispersion surface below. In this way, for some choices of the vector of parameters, one may obtain even a negative band gap (which corresponds effectively to the absence of a true band gap), in case the minimum of the dispersion surface above is smaller than the maximum of the dispersion surface below. Nevertheless, this definition can be preferable to setting the band gap to zero in such a situation, because it allows in principle a gradient-based iterative optimization algorithm to open a band gap, by increasing - iteration after iteration - an initially negative band gap. Otherwise, with the other definition, a zero gradient would be initially obtained, preventing any gradient-based optimization algorithm to increase the initial zero band gap.

The proposed method to solve the optimization problem (1) works as follows.

1) First, a finite number $N$ of admissible choices $\mathbf{x}^{(n)}$ for the vector of parameters $\mathbf{x}$ is generated, by selecting the first $N$ elements of a quasi-Monte Carlo sequence that satisfy all the constraints of the optimization problem (training set). Various choices for the sample size $N$ are made. The final choice $N^\circ$ is the smallest $N$ for which the results of the application of PCA in the successive step 3) are not significantly affected by its further increase.

2) Then, the gradient $\nabla f(\mathbf{x})$ of the objective function is computed, in correspondence of all the $N^\circ$ vectors $\mathbf{x}^{(n)}$ generated in step 1).

3) PCA is applied to the (sampled) gradient field $\left\{\nabla f\left(\mathbf{x}^{(n)}\right)\right\}_{n=1}^{N^\circ}$ of step 2), after removing from each gradient the empirical mean $\sum_{n=1}^{N^\circ} \nabla f\left(\mathbf{x}^{(n)}\right) / N^\circ$ of the gradient field on the whole training set, in such a way to obtain a centered gradient field. The number of principal components kept (or equivalently, the number of principal directions onto which the centered gradient field is projected) is chosen to be equal to the minimal number $p^\circ$ of projections needed to approximate the centered gradient field with a mean squared norm of the reconstruction error $MSRE(p^\circ)$ smaller than or equal to a desired percentage $r$ with respect to the mean squared norm of the centered gradient field.

4) A gradient-based iterative optimization algorithm (namely, sequential linear programming with an adaptively chosen trust region [2]) is used to optimize the objective function, for a given maximum number of iterations $K$. At each iteration of the algorithm, the gradient of the objective function is approximated by its projection onto the (typically $(p^\circ + 1)$-dimensional, otherwise $p^\circ$-dimensional) subspace generated by the empirical mean of the gradient field and by the $p^\circ$ principal directions identified in step 3).

The entire procedure is repeated $N°$ times by choosing, as the initialization of the gradient-based iterative optimization algorithm, one of the $N°$ admissible choices for the vector of parameters made in step 1). Details about such gradient-based iterative optimization algorithm (based on the exact gradient, not on its approximation) are in [2]. It is worth noting that a local decrease in the objective value versus the iteration number $k=1,...,K$ is possible in principle, even when the gradient is not approximated. This occurs when the linear approximation of the objective function in the trust region is poor. To deal with this, the trust region size is changed adaptively, as described in [2]. The largest objective value generated during all the iterations of the various repetitions is produced as output of the proposed method. For comparison purposes, the gradient-based iterative optimization algorithm exploited in step 4) is also applied for other choices of the number $p \neq p°$ of principal directions kept (the case $p+1=8$ refers to the exact gradient of the objective function).

By construction, at each iteration of the gradient-based optimization algorithm associated with an admissible choice for the vector of parameters that does not belong to the boundary of the admissible region of the optimization problem, the approximation of the gradient makes the algorithm update the vector of parameters according to a direction belonging to the subspace identified in step 3), at least if one ignores discretization issues associated with the choice of the trust region (which has the shape of a multi-dimensional box, likewise in [2]). Nevertheless, it is interesting to observe that, in case the current admissible choice for the vector of parameters belongs to the boundary of the admissible region, the update proceeds along a direction determined by projecting the approximated gradient onto the current linearization of that boundary. Hence, although it replaces the exact gradient with an approximation belonging to a subspace having dimension smaller than or equal to $p°+1$, the proposed method makes the gradient-based optimization algorithm potentially able to update the vector of parameters on a subspace having even larger dimension than $p°+1$.

## 3. NUMERICAL RESULTS

The proposed method is applied to the band gap maximization of an acoustic beam-lattice metamaterial having a similar structure as the one investigated in [7]. The number of parameters is equal to $d=8$. Before applying the method, a (standard in PCA) change of variables is made, by mapping the multi-dimensional box associated with the original vector of parameters onto a $d$-dimensional hypercube. Then, the method is applied by considering as the vector $\mathbf{x}$ the resulting transformed vector. The initial choices $N=10,20,30,40,50,60,70,80,90,100$ are made for the number of elements in the training set. The desired upper bound on the mean squared norm of the reconstruction error, expressed as a percentage of the mean squared norm of the centered gradient field, is set to $r=5\%$.

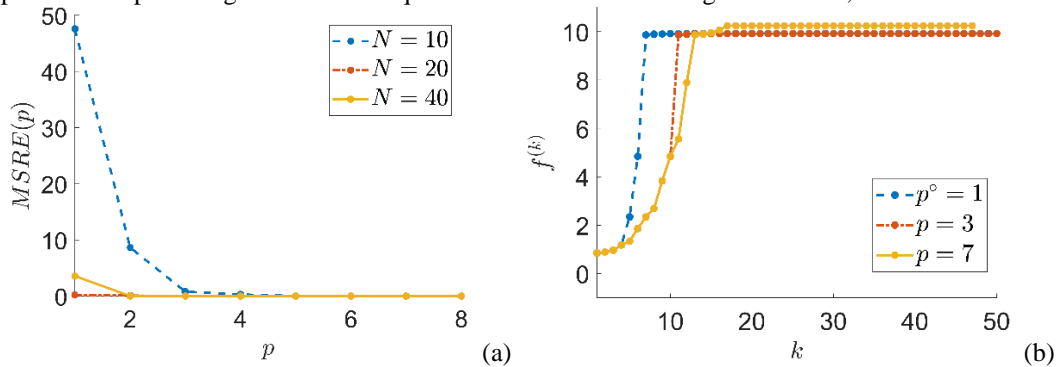

**Fig. 1** *(a) Mean squared norm of the reconstruction error $MSRE(p)$ (in %) of the centered gradient field as a function of the number of principal components $p$ kept, for different sample sizes $N$ and (b) objective value $f^{(k)}$ versus the iteration number $k$ for the best repetition of the proposed method with $p°=1$, and other choices for $p$.*

Fig. 1 (a) shows the mean squared norm of the reconstruction error $MSRE(p)$ of the centered gradient field (expressed as a percentage of the mean squared norm of the centered gradient field) as a function of the number $p$ of principal components kept, for different choices of the training sample size $N$. The graph obtained for $N=30$ overlaps with the one associated with $N=20$. Furthermore, the graphs obtained for $N=50,60,70,80,90,100$ overlap with the one reported for $N=40$. Therefore, this latter value is chosen as the training sample size $N°$. The gradient-based optimization algorithm is applied with a maximum number of iterations $K=200$. The other parameters are chosen in a similar way as in [2]. Fig. 1 (b) shows the band gap $f^{(k)}$ versus the iteration number $k$ for the best repetition of the proposed method and also, for the same repetition, analogous results obtained by selecting other choices for the number $p$ of principal components kept. The figure shows that the proposed method can achieve a performance comparable to the one obtained when the exact gradient is used. Since $p^\circ +1=2$, the gradient approximation takes about $1/4$ of the time required to evaluate the exact gradient.

## 4. CONCLUSIONS

In this work, PCA is applied to achieve dimensionality reduction in a band gap optimization problem for an acoustic metamaterial, by identifying a lower-dimensional subspace of the parameter space and projecting the gradient field of the objective function onto it. The effectiveness of the proposed method is demonstrated by the small dimension of the automatically selected subspace capable to reconstruct the centered gradient field with a given precision and, consequently, by the possibility of reducing the time required for running gradient-based optimization algorithms. This is obtained at the cost of introducing a small approximation error in the gradient, which usually does not prevent one from finding an ascent direction close to the one of the gradient. By selecting a different desired precision in the reconstruction of the centered gradient field, the proposed method could be applied to achieve a different trade-off between the precision in the gradient approximation and the time needed to find it. This is related to recent works on the trade-off between precision of supervision and number of training examples in machine learning problems [8, 9, 10]. A further research direction consists in combining the proposed method with optimization algorithms based on surrogate objective functions constructed by using adaptively chosen sets of training samples [3]. This could be achieved, e.g., by including a new sample in the current interpolation set by moving on the subspace identified by the method. Other possible extensions are the application of PCA to a subset of most recently generated samples of the gradient, and the replacement of PCA with one of its nonlinear versions (e.g., with kernel PCA [11]).

## ACKNOWLEDGEMENTS

A. Bacigalupo and G. Gnecco are members of INdAM. The authors acknowledge financial support from INdAM-GNAMPA (project Trade-off between number of examples and precision in variations of the fixed-effects panel data model), from INdAM-GNFM, from the Compagnia di San Paolo (project MINIERA no. I34I20000380007), and from the University of Trento (project UNMASKED 2020).